# The Art of Generative Narrativity

## Dejan Grba[1] and Vladimir Todorović[2]

*Interdisciplinary Graduate Center, University of the Arts, Belgrade*[1]; *School of Design, University of Western Australia, Perth*[2]
ORCID: https://orcid.org/0000-0002-5154-9699[1]; https://orcid.org/0000-0003-1190-0410[2]
Email: dejan.grba@gmail.com[1]; vladimir.todorovic@gmail.com[2]

| Keywords | Abstract |
|---|---|
| Artificial intelligence<br><br>Computational art<br><br>Generative AI<br><br>Generative art<br><br>Narrativity | Recent advancements in generative artificial intelligence (generative AI) technologies have transformed the computer science discipline of natural language processing. However, generative AI retains the anthropomorphic model of simulating human narrative construction and verbal communication whereas, for artists, the ideational exploration is often more important than human mimicry or even plausibility in storytelling. It sometimes leads to generative experiments with non-verbal forms or events that have the potential to incite narratives through the audience's experience of the works' functionalities, backgrounds, and contexts. In this paper, we focus on such artistic approaches to narrativity. In five central sections, we discuss interrelated exemplars whose conceptual frameworks, methodologies, and other attributes anticipate or underscore the issues of contemporary linguistic automation based on massive datafication and statistical retrospection. In closing sections, we summarize the expressive features of these exemplars and underline their value for critically assessing generative AI's cultural influence and fallouts. |

## 1. Introduction

The integration of large language models (Bommasani et al. 2022) with multimodal (Ramesh et al. 2022) and diffusion models (Yang et al. 2022; Ho et al. 2022) between 2021 and 2022 has led to notable improvements in generative artificial intelligence (generative AI). Large language models show remarkable speed, detail, and extensibility in semantic structuring, concept extraction, and other language-related functions hitherto recognized as hallmarks of human intelligence. Their extensive deployment and rapid commercialization in media generators and chatbots since 2023[1] have made generative AI a part of the mainstream culture, expanded the concerns about the future of storytelling, and raised tensions in pondering narrativity as a human dispositive.² The development and application philosophies of generative AI reaffirm the anthropomorphic programming paradigm that simulates certain aspects of human capabilities (e.g. in composing stories or conversing) with the ultimate goal to emulate them.

---

[1] Popular multimodal media generators are Midjourney, Inc.'s Midjourney, OpenAI's DALL·E and Sora, Stability AI's Stable Diffusion, Meta's Make-A-Video, Runway's Gen-2, Google's Lumiere, and others. Chatbots based on large language models include OpenAI's ChatGPT and Google's Gemini.

Generative methodologies in the arts have a longer history and different priorities. They are media-independent and include diverse approaches to the conscious and intentional interfacing of predefined systems with various unpredictability factors in preparing, producing, or presenting the artwork (Galanter 2016). The expressive value of a generative artwork often depends less on its apparent aesthetics than on its capacity to directly or intuitively engage the audience with the cognitive processes that artists use to devise the intersections of controllable and uncontrollable elements within equally important conceptual and contextual frameworks (Memelink and van der Heide 2023; Grba 2020). These features make generative methodologies useful for conceptual and technical experiments with storytelling and narrativity across disciplines, most notably in literature and computational art, which often transcend the need to simulate human storytelling or even achieve plausibility. Instead, many interesting generative takes on narrativity are configured to nudge the audience toward constructing their own narratives by contemplating the interrelation between the artwork's topic, functional logic, and context.

In this paper, we discuss artistic endeavors that trace or anticipate possible avenues for constructing meaningful narratives besides or beyond the current AI regime of sleek, human-mimetic communication. The advancements in generative AI technologies seemingly change many things in this domain, but key issues of AI narrativity we identified five years ago in *Wandering Machines: Narrativity in Generative Art* (2019) have remained and now become more pronounced, ushering the discursive culture of stochastic parroting (Bender et al. 2021) or ultracrepidarianism (Marcus 2024). Our concept of narrativity combines H. Porter Abbott's "bare minimum" definition of narrative as a "representation of an event or series of events" (2008) with Espen Aarseth's concept of ergodicity (1997)³ and we focus on generative artworks that relate to various literary forms but are not genre-specific or concerned with conventional storytelling. They primarily function as narrative-inciting mechanisms with a certain degree of procedural autonomy, often rely on chance, and require artists' inventiveness to attain a meaningful experiential transference to the audience. Although their hermeneutic space is open and contingent on the spectator's interpretative scope, they frequently require no textual decoding (reading) and unfold as multifaceted non-literary events. These works reveal generative narrativity as a more abstract, non-verbal process of discovery and learning, and we use the term "the art of generative narrativity" in that sense.

---

² For an overview of generative AI's commonly addressed aspects related to artmaking, see Epstein et al. (2023) and Sanchez (2023).
³ Aarseth (1997) defined ergodic literature as literature that requires a reader's nontrivial effort to follow the text.

We trace selected artistic and technical experiments with generative narrativity in five interrelated sections and summarize their common attributes at the end of each. In a section titled "Concealed imperfections," we introduce works that strive to produce plausible narratives by hiding their deficiencies behind a combination of novelty, spectacular presentation, and public fascination with their extraordinary capabilities. The following section, "Unthinking narration," focuses on artworks that take the opposite direction of neither concealing their imperfections nor aiming for narrative plausibility, but instead adopt the functionality of signal-processing machines whose unpredictable outcomes may become narratives if adequately framed. The section "System logic narratives" is about conceptually related works in which the generative logic of narrative systems becomes the main story, often more intriguing than their formal output. Artworks in the section "Reading (into) images" extend this space of creative deconstruction by probing the malleable structural factors that can be used to turn pictures into stories and vice versa. The section "Unbearable lightness of meaning" examines projects that question narrative construction based on statistical retrospection, which generative AI technologies push to new levels of deceptive sophistication and the corporate sector exploits in problematic ways. In two closing sections ("Discussion" and "Conclusion"), we summarize the expressive attributes and critical values of the art of generative narrativity and contrast them with generative AI's human-mimetic processing of language and narrativity.

We intend to show that generative art methodologies expand our emotional and cognitive relationships with narrativity by cultivating proactive anticipation and comprehension of the artworks' logic and context. They also open a more inclusive space for the appraisal of narrativity in the contemporary culture dominated by the anthropocentric generative AI paradigm and tech-biased art-historic myopia. Our critical perspective is informed by ontological changes in modernist avantgardes and postmodernism, which ended the supremacy of aesthetics (principles of nature and appreciation of beauty) in the Western art canon (see Hopkins 2000 and Butler 2003). They transformed the notion of artmaking from the reconfiguration of matter into a cognitive process of relational creativity and discovery, which surpasses the traditional artist-object-spectator hierarchy toward a largely indeterministic meaning construction centered on the spectator's active participation (see Molderings 2010). Thanks to this accentual shift from formal representation to a conceptual exploration that equally favors natural and artificial, physical and imagined elements (Rosen 2022), art appreciation became receptive to objects, events, or processes that do not need to be aesthetically pleasing if their combined features facilitate meaningful communication, discovery, and learning. Therefore, we largely leave the aesthetic analysis of discussed exemplars to the readers' discretion and the future work of authors interested in generative narrativity.

## 2. Concealed imperfections

Amongst a range of 18th and 19th-century mechanical devices that imitated humans and other animals,[4] John Clark's *The Eureka* (1845) was an early predecessor of generative narrative automata. With a pull of a lever, its set of gears, revolving drums, weights, and stop wires arranged a series of lettered staves into Latin words lined up in dactylic hexameter which followed the pattern of adjective, noun, adverb, verb, noun, adjective. This complex machine could randomize words in an estimated 26 million permutations and assemble them into relatively plausible verses (Hall 2017). As in the case of other automata of the time, the public surprise caused by mechanical simulations of traits and activities previously exclusive to living beings was instrumental in concealing many of *The Eureka*'s functional imperfections and logical inconsistencies (Stephens 2023; Hall 2007).

A similarly ambivalent confluence of effects characterized the computational experiments with narrativity throughout the 1950s and 1960s. Early practitioners in that field, mostly engineers with artistic affinities, focused on experiments in extending the syntactic and semantic capabilities of natural language processing (NLP) techniques (D'Ambrosio 2018; Higgins and Kahn 2012). For instance, besides his pioneering work in computer music and games, British scientist Christopher Strachey experimented with computational literature at the University of Manchester. His program *Love Letters* (1952) for a Manchester Mark I computer used a random number generator to combine salutations, nouns, adverbs, adjectives, and verbs from an appropriately compiled lexical database into four-sentence-long love notes signed "M.U.C." (Manchester University Computer). The results are syntactically acceptable and somewhat plausible but semantically inarticulate, resembling the writing of a low-fluency English speaker or the contemporary AI translations of film subtitles (see Sephton n.d. or Montfort 2014). Insinuating the program/computer as an author of awkward love messages was Strachey's tongue-in-cheek reference to his troubled romantic life as a gay man in the 1950s United Kingdom (Hodges 2014; Gaboury 2013). The gesture was not an expression of a belief in computational sentience, and he identified such notions as anthropomorphic projections (Strachey 1954, 25-26).[5] Seven years later, German mathematician Theo Lutz created a stochastic text generator that constructed relatively plausible sentence pairs through a weighted randomization of a 100-word lexicon derived from Franz Kafka's novel *The Castle* (*Das Schloß*, 1926) (ZKM 2020). In this case, however, the output was not signed or otherwise tailored because Lutz was more interested in exploring the quantitative properties of text than in simulating human narratives (Bernhart 2020, 194).

---

[4] Other notable examples include Jacques Vauconson's *Flute Player* (1730s), Jaquet Droz's *Automata* (1768-1774), Wolfgang von Kempelen's *Automaton Chess Player* (1770), and Joseph Faber's *Euphonia* (1845). Many of them relied on artifice (see Paulsen 2020, 4-5; Jay 2001).
[5] Nevertheless, like his university colleague Alan Turing, Strachey did believe that computational emulation of human verbal communication would be possible in the future (Bajohr 2024, 317; Strachey 1954, 31).

Artists-engineers and tech-savvy artists joined in searching for new expressive methods by testing the then-nascent theory of generative grammar and Max Bense's theories of semiotics and techno-aesthetics. For example, Brion Gysin developed the 2,420 lines-long permutational poem *I am that I am* (ca. 1960) with mathematician Ian Somerville who programmed a random generator on a Honeywell Series 200, model 120 computer (D'Ambrosio 2018, 56-58). The *Tape Mark 1* software, created by Nanni Balestrini in 1961, produced generative poetry by recomposing the minimal word units into predefined patterns. As source material, it used the quotes from Lao Tzu's *Tao Te Ching* (4th century BCE), Michihito Hachiya's *Hiroshima Diary* (1955), and *The Mystery of the Elevator* (*Il mistero dell'ascensore*, n.d., attributed to Paul Goldwin by Balestrini). Like Lutz, Balestrini was not interested in simulating human creativity but in exploiting the computational means to quickly resolve certain complex operations on poetic technique (D'Ambrosio 2018, 58-59), and *Tape Mark 1* poems attained semantic plausibility only thanks to his manual editing of punctuation and grammar (Clements 2013; Funkhouser 2007).

NLP experiments continued during the 1960s and 1970s by computer scientists, engineers, and artists of various interests, who worked with stochastic lexicons and syntactic rules to reach plausibility through semantic coherence (Franke 1985). Some of them attempted to use computer graphics techniques for verbal sign manipulation and assembly akin to contemporaneous concrete poetry, which was also influenced by the studies in semiotics (D'Ambrosio 2018). Despite the motivational and methodological variety facilitated by then-powerful computers, these projects resembled the cultural logic of the 18th and 19th centuries' automata. Their public image and reception were swayed by the computer industry's early marketing efforts to leverage art and creativity (Slater 2023).

The conceptual and experiential realms of computational generative narrativity expanded with an early chatbot[6] called ELIZA, developed from 1964 to 1967 and released in 1966 by MIT computer scientist Joseph Weizenbaum to explore human-machine communication. Using symbolic AI techniques for pattern matching and substitution, Wiezenbaum coded several scripts that determined keywords in the textual input, assigned their values, and defined the transformation rules for the (also textual) output (Weizenbaum 1966; see also Wardrip-Fruin 2009, 28-32). The most prominent script, DOCTOR, simulated the conversational technique of the Rogerian school of psychotherapy where a therapist often reformulates patients' statements as questions or repurposes them as prompts for further exchange. Many ELIZA users quickly developed strong emotional involvement with DOCTOR despite recognizing its replies as formulaic and accepted the computer as an intentional discussion partner against their better judgment. Moreover, some practicing psychiatrists at the time seriously believed that DOCTOR could grow into a nearly completely automatic form of psychotherapy, while a belief that ELIZA demonstrated a general solution to the problem of natural language understanding spread in the computer science community (Wiezenbaum 1976, 3-8; see also Wardrip-Fruin 2009, 15, 25-28). Surprised and disturbed by these consequences, Weizenbaum wrote the book *Computer Power and Human Reason: From Judgment to Calculation* (1976) where he described the "ELIZA effect" and set out to show that human-computer-interaction is superficial by default, that anthropomorphizing computers leads to the reduction of humans and other living beings, and that – while human-level AI may be possible – computers will always lack human qualities such as compassion and wisdom and thus should not be tasked with making important choices (see also Berry 2014 and Wardrip-Fruin 2009, 33-34).

---

[6] Chatbots are programs that simulate human conversationalists.

Forty years later, Alexander Galloway (2006) recognized the "ELIZA effect" in some computer games that have the "ability to arrest the [player's] desires in a sort of poetry of the algorithm," which can be leveraged in generative artworks. For instance, Michael Mateas and Andrew Stern's video game *Façade* (2005) uses a chatbot as a core element of the gameplay. Chatting with two virtual characters who are also romantic partners, the player can improve or degrade their relationship which unfolds in a nonlinear story with several predefined branches and endings (Electronic Literature Collection n.d.). While *Façade* conceals its algorithmic limitations and errors to enhance the narrative plausibility, relatively few well-chosen probing inputs stir the program toward selecting a wrong event and reveal the templated repertoire of output options.

Like 18th and 19th-century automata, these technical and artistic explorations of narrativity often owed their success to a variously proportional combination of novelty, spectacular presentation, corporate support, audience's fascination with extraordinary new capabilities, and the skeptics' disbelief in their inventor's claims of genuine autonomy mixed with the admiration for their technical ingenuity.

## 3. Unthinking narration

Unlike artificial narrativity systems configured to produce plausible narratives by hiding their deficiencies or contradictions, many interesting generative artworks do not conceal their limitations and openly adopt the functionality of signal-processing machines. For example, the surrealist technique of automatic writing, which André Breton and Philippe Soupault developed in the early 20th century, calls for writing without thinking, logical reasoning, or consciously manipulating the content. They believed that recording uncontrolled thoughts spawned by memories and subconsciousness provides access to the uniquely deep levels of the mind and that the resulting texts will look edited or censored if we start modifying them with logical reasoning. In theory, automatic writing may be considered an uncensored mind processor. In practice, however, automatically written passages such as:

> The great curtains of the sky draw open. A buzzing protests this hasty departure. Who can run so softly? The names lose their faces. The street becomes a deserted track. (Breton and Soupault 1985) sound too "surrealistically coherent," somewhat like René Magritte's paintings, which were carefully composed for eerie, thought-provoking effects. It is impossible to verify the degree and consistency of the author's ability to unthink the writing process, so the artistic value of automatic writing is as much a matter of fascination and trust as it was for the success of 18th- and 19th-century automata or contemporary AI-powered narrative engines.

Computational generative art abounds in more robust mechanisms that transcode various qualitative and phenomenological aspects of everyday life into forms and situations with new meanings. Once the artist defines how a generative system should work, it is left to perform without additional interventions, often inviting the audience to infer its operational logic. For instance, Ben Rubin and Mark Hansen's installation *Listening Post* (2001-2002) centers around a real-time program that filters text from thousands of online chat rooms, displays it on 200 LED screens, and vocalizes it with a text-to-speech synthesis module in eight audio channels. To determine which message segments will be presented, the program detects simple syntactical patterns in the input text, so it may select only sentences starting with "I am" and output snippets such as "I am 18," "I am from Latvia," or "I am hot!" for some time, then switch to another detection pattern (Bullivant 2005).

With a similar filtering logic, Jonathan Harris and Greg Hochmuth's web project *Network Effect* (2015) generates narratives in a supercut format (MIT Docubase 2015).[7] This work's interface presents a series of keywords for online search and displays an ever-changing stream of videos tagged with the selected keyword. For example, clicking the keyword "kiss" will initiate a torrent of kissing video clips, accompanied by related statistical information (how many people are kissing now, the word frequency for "kiss" in Google Books, etc.). In contrast to its invitingly playful operation, *Network Effect* limits the user's experience to between 6 and 10 minutes per day by factoring in the life expectancy in the country from which it is accessed, which serves as a modern memento mori and a reminder of other, often hidden, constraints and frustrations of our digital lives. Several other generative artworks in the 2010s intersected supercut and automation with uncertainty and arbitrariness to make critical points about surveillance capitalism. For instance, Mimi Cabell and Jason Huff's *American Psycho* (2012) deftly subverted the online profit-motivated recognition of linguistic and behavioral patterns. They mutually Gmailed Bret Easton Ellis' novel *American Psycho* (1991), one page per email, and annotated the original novel text with Google ads injected in each email. They whitened out all original text except chapter titles, placed the ads as footnotes to their (now invisible) trigger words or phrases, and issued the work as a printed book (Muldtofte Olsen 2015; Cabell 2012). Here, interfacing manual data exchange with the whims of Google's AdSense algorithm and its clients' advertising ideas exposes the paroxysms of modern business culture boosted by exploitative datafication.

---

[7] Supercut is an edited set of short media sequences selected according to at least one recognizable criterion. By focusing on specific elements (words, phrases, sounds, scene blockings, visual compositions, shot dynamics, etc.), supercuts accentuate the repetitiveness of forms, routines, and clichés in film, television, literature, music, and online media.

We should note that most artists who elevated supercut from the pastime of meme-obsessed online cultures into a generative art form, such as Christian Marclay, Tracey Moffatt, Jennifer and Kevin McCoy, Marco Brambilla, Virgil Widrich, and Kelly Mark, processed their content material manually. Among them, Dave Dyment stands out by the scale and scope of his projects. For example, to make the supercut video *Watching Night of the Living Dead* (2018), he collected scenes that feature George Romero's film *Night of the Living Dead* (1968) from hundreds of movies and TV shows.[8] He arranged them sequentially along the original film's timeline into a complete zombie classic now featured as the element of cinematic mise-en-scène (Dyment n.d.). While the contents of this and other eminent supercuts were filtered to follow a preset narrative arc, the outcomes are enjoyable for the surprising referentiality of the sourced material. By consciously allowing external, anticipatable but generally unpredictable factors to influence their contents, performance, or atmosphere, these works create new contemplative forms of (fictional) reality.

## 4. System logic narratives

Long before the term "generative art" entered art discourse, the early 20th century avant-gardes were discovering the appeal of generative systems' logic. Artists simultaneously innovated ways to connect with the audience because their relatively simple generative mechanisms produced fragmented or cryptic narratives that required additional layers for the artwork to become fully convincing. Tristan Tzara's *Dada Manifesto of Feeble Love and Bitter Love* (1917) exemplifies this. It is an "algorithm" for making a Dada poem that instructs the reader to take a newspaper article, cut out the words with scissors, put the words in a bag and shake it gently, pick up one word at a time from the bag, and line them up in a sequence following the pick-up order (Tzara 2013). Handling the newspaper, the scissors, and the bag, listening to the sound of paper-slicing, and ordering the cutouts, we slowly enter the magical space in which the assembly procedure, its basic rules, and the way they are presented synergistically connect with the textual outcome and the whole experience can be appreciated for its own merits, usually more significant than simply trying to comprehend the generated text.

---

[8] Because of an attribution error in its title card, *Night of the Living Dead* has always been a public domain film with no licensing or royalty fees and a source of choice for directors who need a cinematic backdrop for their scenes (Hosein 2018).

Dadaist idea of open-ended artwork inspired many artists and art movements after the Second World War. One of them was French poet Raymond Queneau who formed the Oulipo group (Ouvroir de Littérature Potentielle, 1960) around writers and mathematicians interested in generative narrative techniques.[9] Queneau's *A Hundred Thousand Billion Poems* (1961) comprises ten printed and bound sonnets whose verses were horizontally cut into strips that can be flipped over to reveal the underlying verse (ACMI Museum 2024). The work's capability for generating $10^{14}$ different poems invites readers to enjoy its logical properties and creative economy as much as its poetic outcomes, whose plausibility and aesthetics may be impressive but are not paramount. This approach to distributing the expressive "labor" between the artist, their generative system, and the audience strongly resonates with contemporary computational arts. For example, Nick Montfort's *World Clock* (2013) is a 246-page book generated by 169 lines of code, structurally close to Queneau's book *Exercises in Style* (1947) where a single story was written in 99 different styles. In the *World Clock*, the program randomly selects the event's time and place, the protagonist, and the action that initiates an incident and concludes the story in one of 1,440 variations (Montfort 2013). Darius Kazemi, a juror in the computer-generated novel competition NaNoGenMo, jokingly stated that reading *World Clock* is more an exercise in endurance than an indulgence in its literary aesthetics (Dzieza 2014).

A well-contrived generative system can lead to narratives even without linguistic material thanks to the human affinity for establishing mental associations through comparison, abstraction, categorization, analogy-making, and metaphorizing. Nam June Paik's early experiments with sound and video illustrate this potential. Borrowing its concept and title from computer science, his sound installation *Fluxusobjekt Random Access* (1962-1963) elegantly challenges the dictate of linearity in sound reproduction and music. The installation comprised two sets of magnetic audio tapes with the recordings of Paik's musical compositions removed from the reel and cut up into various lengths. One set was assembled on the wall in a stochastic arrangement, another in a parallel grid on a floor-standing looped cloth conveyor. A detached playback head with extended wiring to the amplifier allowed visitors to choose the tape segments and the speed for manually sliding the head over them (Decker-Phillips 2010).

---

[9] Oulipo members' motivation to explore mathematical principles and structures in poetry-making was a close parallel to the early experiments in computational narrativity (D'Ambrosio 2018, 53).

Paik's ethos of hacking and transcoding influenced later, technically sophisticated interactive and generative art projects, such as Ken Feingold's works that use NLP to satirize the dumbness, clumsiness, and heteronomy of modern automation. These chamber-style installations feature idiosyncratically incomplete humanoid robots that aim at establishing highfalutin "existential" conversations with the visitors (*Sinking Feeling*, 2001) or between each other (*If/Then* and *What If*, both 2001). For instance, in *If/Then*, text-to-speech and speech recognition modules in two dummy heads ineptly "strive" for a meaningful dialogue through a mutually triggered vocal synthesis of questions, answers, or comments (Feingold 2021). Marc Böhlen's *Amy and Klara* (2005-2008) uses linguistic transgressions (rudeness and cursing) to critique the corporate AI's normative of gentle, benevolent, and smug speech synthesis (Böhlen 2008). The experience of robotic inanity in these works extends from the uncanny valley awkwardness into a profound sense of absurdity in algorithmic simulations of human exchange. It inspired artworks that exploit the flaws of translation algorithms and coincidentally showcase their increasing accuracy, such as Jonas Eltes' installation *Lost in Computation* (2017) which features a continuous real-time exchange between a Swedish-speaking and an Italian-speaking text chatbot connected through Google Translate service (Eltes 2017).

Cecilie Waagner Falkenstrøm's *Covid-19 AI Battle* (2020) takes the same "dialogical" principle to accentuate the tenuous line between truth and disinformation in an interactive machine learning enactment of contemporary political debates (Waagner Falkenstrøm 2020). It confronts two NLP models with opposing "opinions" on Covid-19: one trained on a dataset of online statements posted by US President Donald Trump and another on statements posted by the head of the World Health Organization, Dr. Tedros Adhanom. Probably the most interesting implication of this work is the analogy between the two learning model's incomprehension of processed linguistic material and the politically fueled "cognitive dissonance" separating the two AI-represented debaters.

The intentional subversion of conventional meaning in these works anticipates the value of deconstructing the human-mimetic "eloquence" and "sleekness" of contemporary large language models, which was presciently illustrated by the spectacle of incongruity and miscommunication in the online "dialogue" between the DOCTOR script of Weizenbaum's ELIZA and Kenneth

Colby's paranoid schizophrenic chatbot PARRY (1972) that Internet pioneer Vint Cerf arranged at the 1972 ICCC conference (Garber 2014).

## 5. Reading (into) images

Throughout the 2000s, artists probed the malleable factors of narrative structuring with sophisticated search and modeling techniques that gained prominence with the "deep learning revolution" happening in that period. For example, in *sCrAmBlEd?HaCkZ!* (2006) Sven König explored the concept of continuous audiovisual synthesis from an arbitrary source pool, which simultaneously destroys and builds narrative structures. *sCrAmBlEd?HaCkZ!* was a program that could extract audio subsamples from stored video material, calculate their spectral signatures and save them in a multidimensional database. Using psychoacoustic techniques, it could search that database in real-time to approximate arbitrary sound input with the closest matching stored audio subsamples that were played in sync with their corresponding video snippets (Van Buskirk 2006). Although *sCrAmBlEd?HaCkZ!* prefigured AI deepfake techniques that emerged a decade later, it has been largely forgotten perhaps because König pitched it to the VJ scene instead of using it to make artworks that could establish intriguing relations between their stored videos and audio input.[10]

Luke DuBois' *A More Perfect Union* (2010-2011) charged NLP with a shrewd sociopolitical critique based on a witty interpretation of the technical term "relational database." It is a map of the United States showing the population's preferred sociocultural identities and mating aspirations. DuBois made a program that sampled 19 million user profiles posted on 21 dating websites and used the associated zip codes to arrange them geographically into 43 maps. The most frequent keywords in dating profiles of local citizens (blonde, cynical, funny, happy, open-minded, lonely, optimist, etc.) replaced the names of their cities, towns, and streets in state and city maps. In federal maps, brightness/saturation ratios represent the relations between female (red) and male (blue) preferences for the most frequent keywords in each state (DuBois 2011).

Taking the opposite direction along the same conceptual axis, Jamie Ryan Kiros' software *Neural-Storyteller* (2015) generates short stories through the semantic analysis of user-uploaded pictures. It identifies forms, objects, actions, and moods in the uploaded images and links them to the keywords and motifs to be processed into narratives by a learning model trained on 11,038 pulp romance novels (Zhu et al. 2015). The model makes unintentionally witty mistakes, e.g. by interpreting a photograph of two clinching sumo wrestlers as a hug of two persons in bikinis. Using a similar methodology, Ross Goodwin's *word.camera* (2015) converts camera-feed images back into short stories. Its Clarifai convolutional neural network extracts tags from input images, which the lexical relations database ConceptNet expands into concepts that a flexible template connects with other concepts and arranges into paragraphs (Merchant 2015). Matt Richardson's *Descriptive Camera* (2012) reverses the transcoding/translation process yet again. Its narrative "picture development" offers a glimpse into the trailblazing data classification labor that the AI industry outsources to underpaid online workers so it can make software products such as these used in the *Neural-Storyteller* and *word.camera*. This device has a conventional digital camera lens and sensor but no display or memory; it sends the captured image to an Amazon Mechanical Turk worker tasked to write down and upload its brief description to the device that prints it out (Richardson 2012).

---

[10] Neither König nor the authors of other similar programs used them to make generative sound-driven video artworks. Mick Grierson's *Study for Film and Audience* (2008) was probably the closest candidate, but it was also conceived as a VJ-style installation (Grierson 2009).

Jake Elwes' *A.I. Interprets A.I.: Interpreting 'Against Interpretation' (Sontag 1966)* (2023) challenges the specter of meaningful narration in generative AI. This 3-channel video processor exploits the mutual input-output feedback between two AI programs.[11] An image-generating diffusion model (Disco Diffusion) is prompted with sentences from Susan Sontag's seminal essay *Against Interpretation* (1966) to synthesize images that are then interpreted back into text by the GPT2 and CLIP image labeling system (Elwes 2023). With a bizarre authoritativeness of the resulting misinterpretations, this work emphasizes the notional reduction of human language to a continuous prediction and chaining of statistically plausible tokens in generative AI. In that sense, it is analogous to earlier works such as Robert Morris' *Self-Portrait (EEG)* (1963) which critiqued the neuroscientific reduction of human consciousness to measurable brain functions (V.A. 1994) and Marc Quinn's *Self-Conscious* (2000) and *DNA Portrait of Sir John Sulston* (2001) which critiqued the reduction of human psychophysiology to genetic code in genomics (Quinn 2001; 2000).

These works' sensibilities unfold as juxtapositions of leitmotifs contained in their input data and generative media outputs (text, sound, image, video). The most arresting aspect of this process is the contemplation of the character of narrative falsity, comparable to the role of levity in surrealists' automatic writing. In both domains, the output may seem frivolous but is also suggestive of the work's backgrounds: the author's obsessions or intentions in automatic writing, data collection mechanisms in social media, and statistical modeling in machine learning media generators.

## 6. The unbearable lightness of meaning

Current generative AI technologies can render elaborate narratives whose meaning is dispersed across syntactically correct, plausible, and even aesthetically pleasing passages that, like the 18th and 19th centuries automata, stir up fascination, concern, and controversy. These technologies leverage sophisticated statistical techniques and powerful computation to enhance one of the essential NLP methods – generating new narratives by analyzing the existing ones – but still exploit the intuitive human ability to select, synthesize, and assess meaning in interpretable constructs. Besides the inherent faults of large language models such as hallucinations,[12] the cogency and overall engagement of generative AI narratives are proportional to their complexity, specificity, and volume. Such seemingly sound, but expressively dubious and relationally feeble narrativity is central to small talk, political speech, journalism, lengthy tech manuals, and verbose science fiction/fantasy sagas. Their common denominator, whether "pragmatic" or gratuitous, is a derivative narrative drift that easily becomes boring and forgetful even if gullible, emotionally susceptible, or ideologically inclined audiences may be willing to absorb, accept, approve, enjoy, or praise it.[13]

---

[11] Elwes used the same method in the *Closed Loop* (2017).

Some generative artworks playfully accentuate the NLP (and literary) notions that all ideas are networks of other ideas and that old stories spawn new ones. For instance, Julius von Bismarck and Benjamin Maus' *Perpetual Storytelling Apparatus* (2008) is a plotter run by software that browses a visual vocabulary of over seven million patents stored in the US government databases and 22 million reference terms to match the word patterns and references between randomly selected patents and print out their accompanying drawings (von Bismarck and Maus 2008). The continuous interweaving of technical innovations' phrases and images in this visual story follows the meandering principle that later computational artworks explored with different programming techniques. For example, poems in Allison Parrish's book *Articulations* (2018, part of the series *Using Electricity*) were generated by extracting linguistic features from over two million lines of public domain poetry to train a learning model, then navigating the fluid paths between the text lines encoded in the model's latent space based on the similarities of their phonetic features (Parrish 2018).

Oscar Sharp and Ross Goodwin's short film *Sunspring* (2016) exemplified the perils of relying on statistics for synthesizing cinematic narratives. Goodwin trained one machine learning model on 162 online-scraped science fiction movie scripts to generate the screenplay and screen directions, and another on a folk songs database to make song lyrics for the soundtrack (Goodwin 2016). Sharp used this material to produce a short live-action film in regular studio settings. Brimming with plot inconsistencies and awkward dialogues, *Sunspring* touches upon several issues of its underlying cultures. Compared with human-written science fiction narratives, its incongruity offers an analogy for the frivolity or nonsensicality of science fiction imaginaries. Like the Dadaist poetry, its engagement grip depends on knowing the production context but, more importantly, Goodwin and Sharp's satirical application of AI to filmmaking parodies Hollywood's trademark regurgitation of themes from earlier films and anticipates the current use of machine learning for screenplay analysis and design (Grba 2017, 390-392).

---

[12] Hallucination is a phenomenon in which the learning model outputs – with formal linguistic fluency that feigns self-confidence – content that is false, incorrect, nonsensical, or in other ways inconsistent with real-world facts or user inputs (see Xu et al. 2024 and Huang et al. 2023).

[13] Interventions such as the Sokal hoax in 1996 (Sokal and Bricmont 1998) and its programmatic offspring the *Postmodernism Generator* (Bulhak 1996) unmasked derivative proliferation of narratives within the "science wars" arena of the 1990s but are equally relevant today (see Sokal 2008).

Like commercial cinema, complex entities such as governments, industry, marketing, finance, insurance, media, and advertising involve frequent information exchange and processing, which can be controlled by automated quantization and data collection, behavioral tracking, predictive modeling, and decision-making manipulation. The AI industry couples massive digital datafication with intricate algorithms to increase the extent and efficacy of these strategies as well as their undesirable effects which arise from the disparities between business priorities in maximizing wealth and competitive power, the impact of AI products on various demographic groups, and broader societal interests (Zuboff 2019; O'Neil 2016). For instance, social media platforms convert users' input into a generative fuel of corporate data narratives, both explicit (search inputs, clicks, selfies, blogs, news updates) and implicit (behavior patterns, intentions, desires, psych profiles). Paolo Cirio and Alessandro Ludovico's hacking action *Face to Facebook* (2010-2011) brilliantly repurposed techniques for pattern recognition, NLP, and computer vision over the established online protocols to make strong critical points about this AI-powered data-narrative transfer pipeline. The artists wrote a program that exploited Facebook's security vulnerability to navigate one million user profiles, randomly download 250,000 names, profile pictures, locations, and likes, and process the profile pictures with a custom face recognition software that classified gender (male/female) and "temperament" into six categories: climber, easygoing, funny, mild, sly, and smug. They added the classification results to the gathered users' data and published them on a fake dating website called Lovely-Faces.com. During a week while the site was online, it had 964,477 views and the artists received several letters from Facebook's lawyers[14] and a variety of visitors' reactions: from requests to be removed (which were diligently satisfied), through lawsuit warnings and death threats, to commercial partnership proposals (Cirio and Ludovico 2011). By exposing the centrality of our participatory-exploitative adherence to social media and other online services, *Face to Facebook* suggests that our complacency, narcissism, or ignorance help us fall for sinister corporate agendas (Grba 2020, 73) while our fetishization of privacy protects us from acknowledging that the stories *of* us (as told by the data we provide) may be more colorful than the stories *we tell* about ourselves (Rosenberg 2018).

---

[14] The correspondence between the artists' lawyer and Facebook's legal team is available at www.face-to-facebook.net/legal.php.

These artworks provide unorthodox and seemingly counterintuitive ways to encounter the various backgrounds of AI's evasive meaning-making mechanisms and expand our critical views on AI-automated narrativity and discursive culture in general. They bypass conventional storytelling methods, but their outcomes can conjure rich narratives implicitly by inviting, motivating, or provoking the audience to think about artworks' generative processes within wider contexts.

## 7. Discussion

A common aspect of most exemplars we described is an affirmation of the active roles and responsibilities shared between the artist and the audience for actualizing the work. Successful generative artworks require our feeling of presence, discovery, examination, and evaluation of our sense of meaning to help us blend perceptual matrices into new narrative structures that relate to our being in the world. More importantly than telling stories, they stimulate our imagination and motivate creativity by suggesting or revealing their makers' thinking processes that challenge our notions in engaging ways. The joy and fun in experiencing generative artworks come from their ability to mobilize our mental resources for building anticipations, concepts, and knowledge, which can incite surprising insights and emotions (Grba 2015). By reiterating the simple question: "What *is* a narrative?," generative artworks inspire our amazement with storytelling and broaden our self-critical understanding of narrativity in the sense that we make our stories, and they make us in return. By elevating the storytelling dynamics beyond conventional representation, they also stir our appreciation that a narrative is always uniquely performative, a story always an unfolding event flavored with unpredictability.

Although many discussed artworks invite the audience to enter the tricky space of authorship assignment ranging from causal (only human) through distributed to the autonomous or strong artificial,[15] they ultimately reaffirm that our self-awareness and its embeddedness in physical and social realities inform the meaning and appeal of our narratives. Explicitly or implicitly, they also call attention to the sophistication and shortcomings[16] of generative AI narratives and indicate their roles in social engineering, economic exploitation, and political manipulation. By churning out semantically plausible but not necessarily sensible discursive episodes, some of these artworks simultaneously criticize and unwittingly exacerbate the AI-powered outpouring that inflates the already fascinating abundance of human-made linguistic nonsense. This may remind us that we are neither innocent nor sincere parties in this arrangement. Our self-serving/advantage-seeking strategies in adopting and using AI products (un)willingly uphold their socioeconomic authority and encourage their owners to exploit our desires further.

---

[15] For various perspectives on authorship in generative AI studies see Bajohr (2024), Grba (2024, 8-9), and Wilde et al. (2023).

[16] Large language models simulate certain manifest features of human verbal abilities but do not embody all their relevant aspects due to technical difficulties, insufficient knowledge about human cognition, high level of operational heteronomy, and isolation from the human existential realm. See Mitchell and Krakauer (2023), Bender et al. (2021), and Mitchell (2019).

However, straightforward artistic challenges of generative AI, such as Elwes' *A.I. Interprets A.I.* discussed in section 5, remain atypical among largely cynical, softly critical, ambiguous, or counter-effective works. They often use vogue techniques and "fringe" post-digital aesthetics but reshuffle AI tropes that are already widely familiar or trivial. For example, Silvia Dal Dosso's *The Future Ahead Will Be Weird AF* (2023) is a Corecore-styled mashup[17] of Internet-sourced deepfake videos narrated by the AI-synthesized voice of Adam Curtis reading the artist's essay about climate change, deepfakes, generative AI, and post-truth (Dal Dosso 2023). Jonas Lund's *The Future of Something* (2023) is an omnibus of seven AI-generated video vignettes that parody different types of support group discussions about AI-related anxieties, primarily the fear of human displacement by automation and the widespread complacency to the current AI regime (Lund 2023b). Similarly, Lund's video *The Future of Nothing* (2023) comprises short slideshows of art-themed AI-generated images accompanied by diary-style narrations in AI-synthesized voices about the consequences of automation for human roles in art and creative industries (Lund 2023a).

Other artworks vacillate between criticality and self-indulgent gimmickry or inadvertently become counter-effective. For instance, Bjørn Karmann's *Paragraphica* (2023) is a physical device prototype that uses web APIs to collect the address, weather, time, and nearby places data at the location it is used, which is then inserted in a prompt template for the Stable Diffusion to generate the output image (Karmann 2023). Without the artist's brief online note about the aim to allow the unconventional, AI-mediated perception of the world, it is hard not to see *Paragraphica* as a pitch for a nifty consumer gadget that feeds situational data to a popular text-to-image service. As another example, Jan Zuiderveld's *Conversations Beyond the Ordinary* (2024) reinforces anthropomorphism by attempting to critique it. It is an interactive installation with three office appliances (a coffee vending machine, a photocopier, and a microwave) whose AI-powered "behavioral quirks" and unusual "personalities" "foster empathy and humor in existential inquiry and sociopolitical commentary" and "invite reflections on power, agency, and creativity" (Ars Electronica 2024). The microwave's vocal interface expounds a critique of society's dependence on technology and the erosion of human agency in the quest for convenience. The photocopier scans visitors' drawings as templates to generate and print new images. The coffee machine's vocal interface "laments" on its repetitive existence and users' disrespect, and visitors must vocalize sincere interest in the device to make it brew them a cup. Here, the "smarty-pants" bending and linguistic fluency of large language models coerce visitors' "empathy" but nothing in the hacked devices' functionality (or in the project's online documentation) distinguishes between their intentionally anthropomorphized features and the real harms of unintentionally anthropomorphizing AI systems.

---

[17] Corecore or CoreCore is the Internet youth culture aesthetic aiming to capture the post-2020 *zeitgeist* by mashing up video clips over emotional music soundtracks (see Wikipedia 2024).

This acquiescent criticality may be a consequence of generative AI's sociocultural reign (see Grba 2024). Against the expressive variability of generative methodologies we traversed throughout this paper, the hyped-up trends in AI tech push the online, media, and pundit vocabularies and views toward historically ignorant exclusivity, and generative art is one of its casualties. Even before "generative art" became the term mainly associated with AI-generated artefacts, it had been conflated with earlier computational art practices that involved randomness, complexity, and machine learning (Benney and Kistler 2023). This terminological "roaming" illustrates how foregrounding the artistic uses of faddish technologies in popular and professional discourse supplants the appreciation of diverse, historically founded art fields with marketing labels and sanctions their uncritical appreciation. At the same time, generative AI systems impose into the creative process an opaque conglomerate of models, prediction algorithms, and filters as a source of uncertainty whose parameters are effectively unknowable[18] in contrast to the canonical generative artwork architecture where all elements have a reasonable degree of cognitive resolvability. These limiting factors push the notion of artistic generativity toward the realm of passivized consumerist authorship where narrative meaning and expressive persuasiveness dissolve in the complexity of AI technologies, dubious aspirations of their owners, and bewilderment of their users.

## 8. Conclusion

While generative AI's progressively improved simulation of human narration and linguistic abilities is useful in certain application scenarios, it also fosters an ever-elaborate façade of sleek, "charmingly" human-like but ultimately unreliable communication that veils the technical problems of large learning models, broader conceptual issues in AI science, and the ideological tendencies of the AI industry. These contradictions are paralleled by big AI companies' clashes over closed- vs. open-source policies, the continuous release of insufficiently tested products in a race for market supremacy, and extensive lobbying against public oversight and government regulation. One of the AI's central problems, besides its roots in the military-industrial complex and close alignment with capitalist political interests, is a paradox of technically elaborate and materially exhaustive attempts at emulating human intelligence despite the insufficient understanding of human intelligence based on formally robust definitions and rules (Larson 2021; Mindell 2015). Abiding by the old computationalist premise that intelligence will "spontaneously" emerge from its ever more intricate simulations (Bender et al. 2021), the AI industry burrows further into this paradox without resolving it. Instead, it sustains an illusion of the coherence of its underlying concepts and the indispensability of its products by delivering means that churn out elusive "food" for our evolved tendency toward extracting meaning from formally correct but not necessarily meaningful patterns.

---

[18] Elwes' *A.I. Interprets A.I.* critiques this arbitrariness in both artistic and wider sociopolitical contexts.

Generative AI's human-mimetic approach to narrativity is not particularly interesting from artistic viewpoints that seek inventiveness in revealing new facets of nature and reality instead of achieving virtuosity in copying their apparent aspects. This approach may be pragmatic and lucrative, but it continues a dangerous trend of translating fallacies, biases, stereotypes, and tainted political ideas from insufficiently tested algorithms and arbitrarily collected data into widely deployed systems whose operation often has profound sociopolitical consequences. The anthropomorphic AI paradigm also normalizes an undeserved sympathy for non-living entities and diverts researchers from alternative pathways that could offer novel insights into the processes and social functioning of the human mind. In contrast, the creative embrace of context, connotations, and contingencies in generative art relates to the meaning of the Latin noun *error* "wandering about" and encourages deviations from the regular, usual, or expected modes of thinking and expression. It traces the routes for addressing generative AI narrativity by challenging its apparition of fluency, benevolence, and trustworthiness with interventions against content legibility, determinacy, and trivial interpretation.

## Bionotes

Dejan Grba is an artist and scholar who explores the expressive and relational features of emerging media arts. His work has been exhibited and published worldwide. He has served as an associate professor in the Digital Art Program of the Interdisciplinary Graduate Center at the University of the Arts in Belgrade, and as a founding chair of the New Media department at the Faculty of Fine Arts in Belgrade. He was a Visiting Associate Professor in the School of Art Design and Media at the NTU in Singapore.

Vladimir Todorović is an artist, researcher, and educator living in the Whadjuk region, Australia, where he is the chair of Fine Arts in the School of Design at UWA. He works with new technologies for immersive and generative storytelling. His art explores our relationship with the changing environment and our misuse of past, current, and future technologies. His projects have won multiple awards and have been showcased at festivals, exhibitions, museums and galleries across the globe.